# Calibration-free quantitative phase imaging using data-driven aberration modeling

TAEAN CHANG,[1,2] YOUNGJU JO,[1,2,3,4] GUNHO CHOI,[3] DONGHUN RYU,[1,2] HYUN-SEOK MIN,[3] AND YONGKEUN PARK[1,2,3]*

[1] *Department of Physics, Korea Advanced Institute of Science and Technology (KAIST), Daejeon 34141, Republic of Korea*
[2] *KAIST Institute for Health Science and Technology, Daejeon 34141, Republic of Korea*
[3] *Tomocube Inc., Daejeon 34051, Republic of Korea*
[4] *Present address: Department of Applied Physics, Stanford University, Stanford, California 94305, USA*
*\*yk.park@kaist.ac.kr*

**Abstract:** We present a data-driven approach to compensate for optical aberration in calibration-free quantitative phase imaging (QPI). Unlike existing methods that require additional measurements or a background region to correct aberrations, we exploit deep learning techniques to model the physics of aberration in an imaging system. We demonstrate the generation of a single-shot aberration-corrected field image by using a U-net-based deep neural network that learns a translation between an optical field with aberrations and an aberration-corrected field. The high fidelity of our method is demonstrated on 2D and 3D QPI measurements of various confluent eukaryotic cells, benchmarking against the conventional method using background subtractions.

## 1. Introduction

Quantitative phase imaging (QPI) has rapidly emerged as a prominent imaging modality in life sciences and medicine owing to its non-invasive, label-free, and quantitative visualization at the individual single-cell level [1]. In general, QPI systems are implemented using laser interferometry to measure optical phase delay induced by a transparent specimen, although several non-interferometric QPI techniques have also been investigated [2-5]. Due to its label-free and quantitative imaging capability, QPI has been widely used for various physical and biological studies, including cell biology [6, 7], infectious disease [8, 9], microbiology [10, 11], and therapeutics [12].

One of the important issues which can affect and deteriorate QPI imaging quality is aberration and coherent noise, resulted from the use of a coherent illumination source. Ideally, plane wave illumination that impinges on a sample is required for the precise measurement of the sample-induced optical phase delay [Fig. 1(a)]. However, in practice, the presence of optical aberration induces a deviation in the measured phase image from the correct one [Fig. 1(b)].

To address this issue, several approaches have been reported and utilized. Although the coherent noise might be alleviated by employing spatially or temporally incoherent illumination [13-18], the resulting short coherence length can degrade the performance of QPI and complicate its optical system. Other straightforward methods to circumvent optical aberration in QPI are background subtraction [19] or sample translation [20]. In the background subtraction method [Fig. 1(c)], optical aberration is removed by subtracting the measured phase image of a sample from that of a background area without objects [Fig. 1(c)]. This background subtraction method is straightforward, but it requires additional measurement of an object-free background and is not always possible, particularly for confluent samples, including biological cell aggregates and tissue slices.

The sample translation method retrieves and removes optical aberration from two differential images with slight lateral shifts [20]. Although the sample translation method is not restricted to the confluency of samples, it still has a strict restriction that samples must not change during the shifts, which prevents probing the fast dynamics of biological phenomena. Alternatively, fitting with Zernike polynomials was reported [21]. Even though it does not require additional measurement, this fitting method only works for low-order aberration because of its assumption.

Here, we present deep-learning-based aberration compensation for calibration-free QPI. Using a paired dataset (optical field with aberration and aberration-corrected field using the background subtraction), we trained a deep convolutional neural network (DCNN) that learns the aberration patterns of an optical system and suppresses the aberration of input optical field images [Fig. 1(d)]. The proposed network, inspired by U-net [22], used an encoder-decoder network with skip connection in order to circumvent the bottleneck for information (Fig. 1e; Appendix A) [23, 24]. Our method enables the single-shot aberration-corrected measurement of an optical field. The main advantage of our approach is that any dynamic or confluent samples can be obtained via QPI because it does not require any additional images or assumptions. To demonstrate the applicability, we performed 2D and 3D QPI of various samples, including polystyrene microspheres and eukaryotic cells. We also verify the high-fidelity performance of the present method in comparison to the background subtraction method.

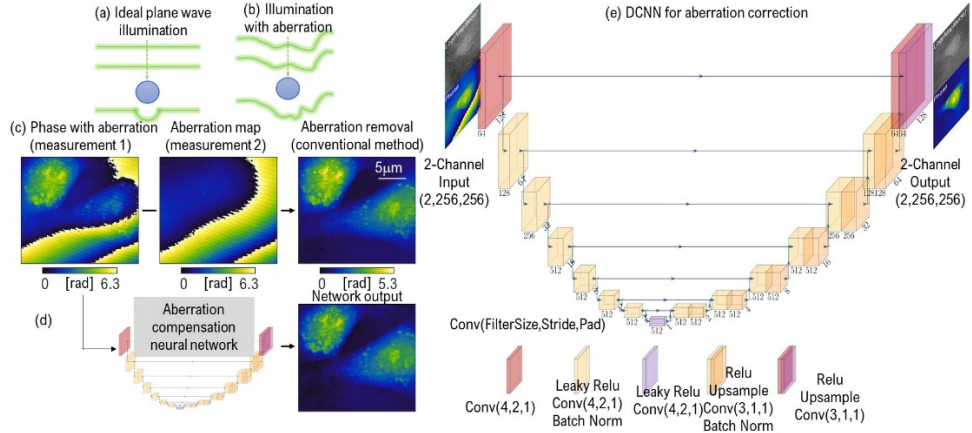

Fig. 1. (a)–(b) The distortion of an illumination beam is caused by imperfections of optical systems, which should be removed to image clearly in holographic measurements. (c) The conventional background subtraction method requires additional measurement to compensate for aberration. (d) The proposed aberration compensation neural network can generate an aberration-compensated phase image directly from a distorted phase without additional measurement. (e) Details of the network architecture. Conv: 2D convolutional operation. Leaky Relu: Leaky rectified linear unit. Batch Norm: 2D batch normalization.

## 2. Results

### 2.1 Calibration-free 2D Quantitative Phase Imaging

For experimental verification, we tested our method with eukaryotic cells (HeLa, NIH-3T3, HEK-293T, MDA-231, and COS7 cells). The representative results of NIH-3T3 and Hek-293T are shown in Fig. 2. The input, output, and ground truth (GT) complex field images of an NIH-3T3 cell, and confluent HEK-293T cells are presented in Figs. 2(a) and 2(b), respectively. The aberrations in the optical input fields are effectively suppressed in the network outputs, and the results are compatible with those obtained using the conventional background subtraction method (GT in Fig. 2).

Parasitic fringe patterns, caused by multiple reflections (red arrows in Fig. 2), are suppressed. Intriguingly, the phase discontinuity caused by phase wrapping in the input phase images (the white arrows in Fig. 2) is also clearly removed in the network output [Fig. 2(a)] This implies that through the training of a paired data set (input and GT images), the algorithm automatically learns the information about an ideal imaging transfer condition, suppressing any deviation from an ideal optical imaging condition.

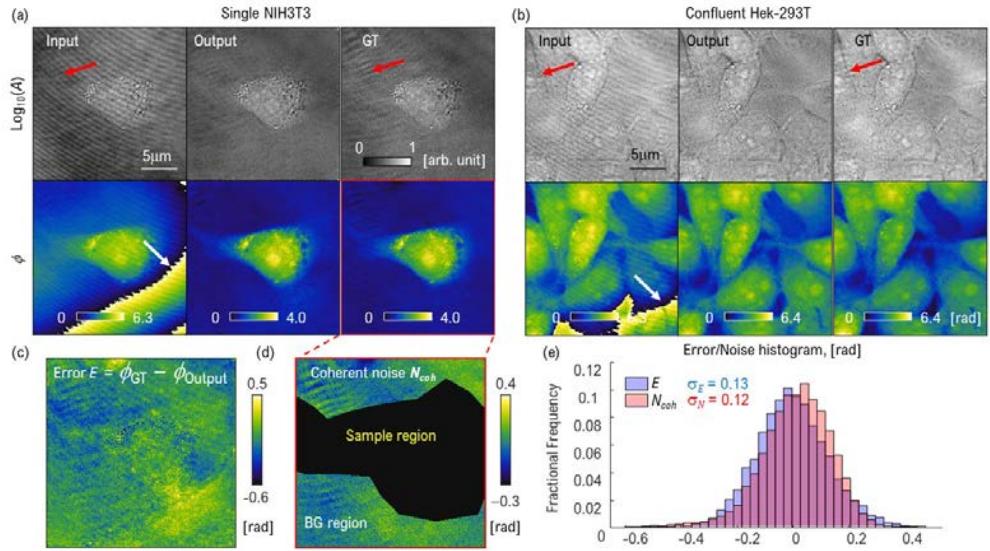

Fig. 2. Experimental results. (a)–(b) Complex field images of an NIH-3T3 cell and confluent Hek-293T cells, respectively. The input fields embracing aberration and phase wrapping (left column), our deep learning results (middle column), and their GTs (right column) that are obtained from the background subtraction method are visualized. (c) The phase error map ($\phi_{GT}$-$\phi_{Output}$) of the image in (a) is presented. (d) In order to compare the level of the phase error with the coherent noise of the setup, the background region (no-sample

region) in the image (a) is selected. (e) The Fractional histogram of the phase error E and coherent noise $N_{coh}$ are presented with the standard deviations separately.

The capability of our method was further validated with HEK-293T cells that densely occupy the entire field of view (FoV) [Fig. 2(b) and Appendix D]. Similar to the NIH-3T3 case, the fringe noises and phase ambiguities were removed successfully, generating sufficiently clean images for probing biological information. This result indicates that our method can be effectively applied to confluent samples where no background region can be obtained. In such a scenario, on the other hand, it is difficult or even impossible to use the aberration correction methods that must exploit a background. We compared the difference between the GT and the network output image [Fig. 2(c)]. The cross-correlation and root mean square error (RMSE) were measured as 0.982 and 0.127 rad [NIH-3T3 cell, Fig. 2(a)], and 0.950 and 0.317 rad [HEK-293T cells, Fig. 2(b)], respectively. While negligibly blurred, the subcellular features such as nucleolus and vesicles are preserved well in the network output.

The network error level is compared with the coherent noise level of the optical system used to validate the proposed method quantitatively. The absolute phase error map $E$ between the present method and the conventional background subtraction method was calculated as $E = \phi_{GT} - \phi_{Output}$, where $\phi_{GT}$ and $\phi_{Output}$ are the phase images of GT and the network output, respectively. The phase error map for the NIH-3T3 cell is shown in Fig. 2(c). The background region was obtained from the measured FoV by manually masking out the cell to compare the phase error with the coherent noise level [Fig. 2(d)].

In Fig. 2(e) the fractional histogram of the phase error E (blue) and coherent noise $N_{coh}$ in the background region (red) are presented with their standard deviations. Both centered at zero, the standard deviations (0.13 rad and 0.12 rad) are comparable, indicating that the performance of our approch achieves the fundamental upper limit. Similar quantifications for the entire dataset can be found in Appendix E: the mean cross-correlation is 0.93, the mean RMSE is 0.23 rad.

## 2.2 Calibration-free 3D Quantitative Phase Imaging

The proposed method can be further utilized in 3D imaging. A three-dimensional refractive index (RI) tomogram is reconstructed from multiple 2D optical fields at various angle illuminations using the principle of inverse light scattering, which is known as optical diffraction tomography (ODT) [Fig 3(a)]. ODT is a 3D QPI technique, and it reconstructs the 3D RI tomogram of a sample from multiple 2D optical field images by inversely solving Helmholtz equation [25].

In ODT, it is essential to correct each optical field's aberration because varying aberration in the angle-scanned optical fields, which frequently occur in practice, can deteriorate the image quality of reconstructed tomograms [Fig 3(b)]. The green dashed arrow indicates the significant noise caused by aberrations.

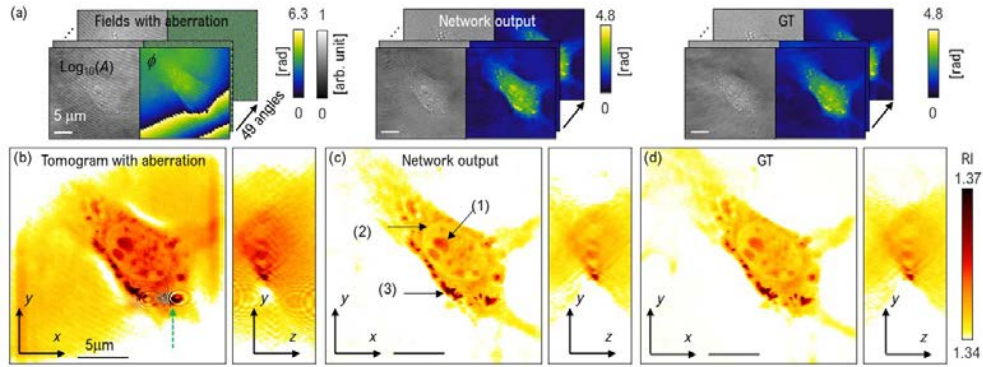

Fig. 3. Experimental results of diffraction tomography. (a) The optical field images of an NIH-3T3 cell obtained from 49 angle-scanning illuminations that suffer from aberrations and phase wrapping (left). The aberrations in the field images are removed, and necessary processes including phase unwrapping are conducted together by our proposed method (middle). The optical field images obtained from the background subtraction method for validation (right). (b) The tomogram of the NIH-3T3 cell that is reconstructed using the optical fields with aberrations. (c) The tomogram of the NIH-3T3 cell that is reconstructed from the aberration-removed fields in (a) using the proposed method. (d) The GT tomogram of the NIH-3T3 cell is reconstructed using the background subtraction method. The solid black arrows indicate the organelles, including nucleoli (1), nuclear membrane (2), and lipid droplets (3). The green dashed arrow indicates the significant noises induced by aberrations.

Figure 3(c)−3(d) show how our method can significantly improve the performance of ODT, benchmarking against the background subtraction method. The aberration correction on 49 optical fields via our method leads to the substantially improved reconstruction of the tomogram, compared to the reconstructed tomogram from the optical fields with aberration. The subcellular structures of NIH-3T3 were clearly restored in our network output. The solid black arrows indicate that the morphology of organelles, including nucleoli (1), nuclear membrane (2), and lipid droplets (3), were successfully reconstructed, as done in the reconstructed tomogram using the background subtraction method (GT).

The results show that this method improves the applicability of QPI in two ways. First, the proposed method does not require any additional measurements, enabling "single"-shot QPI. This aspect offers a robust phase measurement without sample restrictions such as staticity or sparsity. Second, our network can perform not only aberration correction but also a robust phase unwrapping process, which is difficult to conduct owing to its ill-posedness. The presented results reconfirm the use of deep learning for phase unwrapping, which has previously been reported in Ref. [26, 27].

## 3. Conclusion

In sum, we proposed and experimentally demonstrated a deep-learning approach that performs aberration correction. The deep-learning model learned mapping from the 'input' optical field with aberrations to 'output' aberration-corrected optical field using the background subtraction method. The model, trained with the paired dataset, generates aberration-corrected optical fields without any additional measurements, enabling single-shot QPI. We have verified the robust performance of our method by successfully operating 2D QPI on various eukaryotic cells and optical diffraction tomography where the 3D refractive index of a biological sample is reconstructed from the 2D aberration-corrected optical fields. Recently, deep learning has been applied to segment background region in the measured FoV [28]; however, it necessitates object-free background areas in the image to model aberrations based on the Zernike functions, which is fundamentally different from our method. We believe our approach, capable of imaging dynamic and confluent samples, can be used to investigate perplexing biomedical phenomena in diverse disciplines.

Several subsequent studies can be conducted to enhance the proposed method. First, advanced network architectures, such as Dual-CNN or LS-DNN [29, 30], may mitigate some blurring and pixelization artifacts that have appeared in our results. Next, a more diverse dataset can further generalize our method. As we used a single QPI instrument and six types of eukaryotic cells in our work, our framework may suffer from generalization when dealing with unseen data distribution or different imaging systems. The data preparation process can also be simplified and automated by manufacturing permanent reference samples with various RI distributions. [31]. Also, the present approach can be sequentially used or combined with a deep-learning-based phase retrieval algorithm [32-34], which can potentially even expedite the whole imaging and reconstruction process. Finally, artificial-intelligence-aided QPI approaches can be further utilized with the proposed method owing to the high throughput and automation [35].

## Appendix A: Network architecture for aberration compensation

The encoder extracts the features of the input image at various scales through the successive down-sampling convolutional blocks. The input image is first convolved with $4 \times 4$ filters with stride 2 and pad 1 to create 64 feature maps. Down-sampling blocks are successively applied, extracting feature maps of different levels. The blocks consist of leaky rectified linear units (Leaky Relu) [36], $4 \times 4$ convolutions with stride 2 and pad 1, and batch normalization [37], typically used to reduce internal covariate shift and training time. The number of convolutional filters doubles with each layer until it becomes 512 while the image size halves. The number of layers is 8 in this case (the input image size = 28) to have features of $1\times1$ dimension at the end of the encoder. The decoder reconstructs the output image of the same dimensions with the input from the acquired features of the encoder. The up-sampling blocks consist of ReLu, bilinear interpolation (factor of 2), $3 \times 3$ convolutions with stride 1 and pad 1, and batch normalization. Magnification with convolution was used instead of transposed convolution to avoid the checkerboard artifact [38]. The dimensions of blocks are symmetric with respect to the encoder, except that features of the same dimension in the encoding path are concatenated to preserve spatial information. The last blocks of the encoder and decoder do not contain the batch normalization.

## Appendix B: Loss function

Among various metrics to compare two images, we utilized $l_1$ norm [Eq. (1)] and SSIM (contrast component only) [Eq. (2)] [39]. We only exploited the contrast component because the global phase, which contributes to the offset, does not have any physical meaning. For network $G$, a network input $x$, and ground truth (GT) $y$,

$$\boldsymbol{Loss_{L1}(G) = E_{x,y}[\|y - G(x)\|_1]}, \tag{1}$$

$$\boldsymbol{Loss_{SSIM}(G) = E_{x,y}[1 - SSIM(G(x), y)]},$$

$$\boldsymbol{SSIM(x, y) = \left(\frac{2\sigma_x \sigma_y + c}{\sigma_x^2 + \sigma_y^2 + c}\right)^\alpha}, \tag{2}$$

where $E[X]$ is the expectation of the random variable $X$, $\|x\|_l$ is the $l_1$ norm of a vector $x$, and $\sigma^2$ is the local variance. Because minimization of norm generated blurry images, we combined the contrast component of SSIM [Eq. (3)] in order to achieve higher resolution:

$$Loss = Loss_{L1} + \lambda Loss_{SSIM}, \qquad (3)$$

where we used $\lambda = 1$.

When the loss function consisted of SSIM only, the optimization process was highly unstable, failing frequently. The performance of each loss combination was tested by calculating the average field cross-correlation error (FCE) [Eq. (4)] and phase RMSE [Eq. (5)]. For a network output and GT complex electric field $E = Ae^{i\phi}$,

$$FCE = |E_{GT}^* E_{Output}|, \qquad (4)$$

$$RMSE_\phi = \sqrt{E[(\phi_{GT} - \phi_{Output})^2]}. \qquad (5)$$

The linear combination of $l_1$ norm and SSIM achieved the lowest error, which was chosen for the present work.

**Appendix C: Dataset and training**

To obtain the dataset, the optical field images of biological cells in various cell lines (HeLa, NIH-3T3, HEK-293T, MDA-231, and COS7 cells) were measured using a commercial 3D QPI instrument (HT-2H, Tomocube Inc., Republic of Korea). The dataset was prepared by pairing the optical fields with aberrations and fields with aberration correction using the background subtraction method. To exploit the amplitude-phase correlation, the network was designed to handle 2-channel (amplitude and phase channel) images for input and output. For the ground truth training dataset, phase images were retrieved using a phase retrieval algorithm [40, 41], and the $2\pi$ ambiguities in phase maps were resolved using a phase unwrapping algorithm, namely Goldstein's algorithm [42].

During the training process of the algorithm, the network parameters were updated by gradient-descent optimization, minimizing the loss function consisting of structural similarity index map and *L1* norm [43]. We used a desktop computer with a CPU (XeonTM CPU E5-2630, Intel), and one graphic processing unit (GeForce GTX 1080Ti, 11 GB internal memory, NVIDIA) for training and testing. Once the training is completed, the process of converting an image with aberration into an aberration-corrected image takes 0.8 ms (with a batch size 100 and 15 ms with a single-image batch). All optical field images were cropped to 256 ×256 pixels before entering the network.

The learning curve of our network is depicted in Fig. 4. The average loss was calculated using the training set (11809 images) and validation set (11858 images) at each epoch during the training. Although our network was nearly optimized within 24 hours of training and validation, we decided to train the network for sufficient time (approximately 240 hours) to fully learn the image-to-image translation. We stopped the training when the validation loss achieved 28.53 at 2223 epochs.

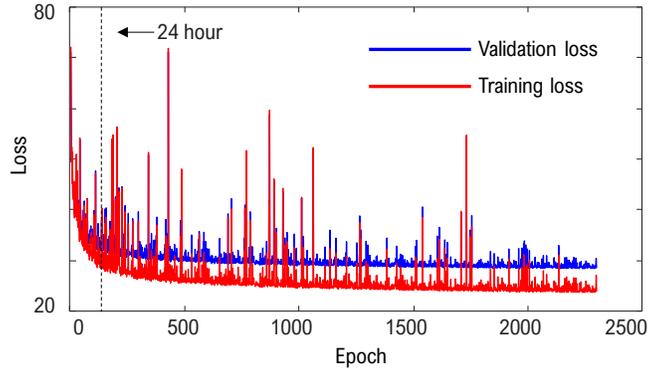

Fig. 4. The learning curve of the training of the DCNN. The model can be optimally trained within 24 hours.

**Appendix D: Demonstration on microbeads and various eukaryotic cells**

To validate that the proposed network did not learn the specific sample distributions in the training dataset, but the ensemble of aberration patterns of the optical system used, we also imaged known polystyrene (PS) microsphere (Fig. 5). The phase images of the PS microsphere are presented in Fig. 5(a). The input phase image containing aberration and phase wrapping (left), the network output (middle), and the GT (right) obtained from the BGSM are visualized. Phase delay along the red dashed lines in the insets of Fig. 5(a) are also plotted in Fig. 5(b).

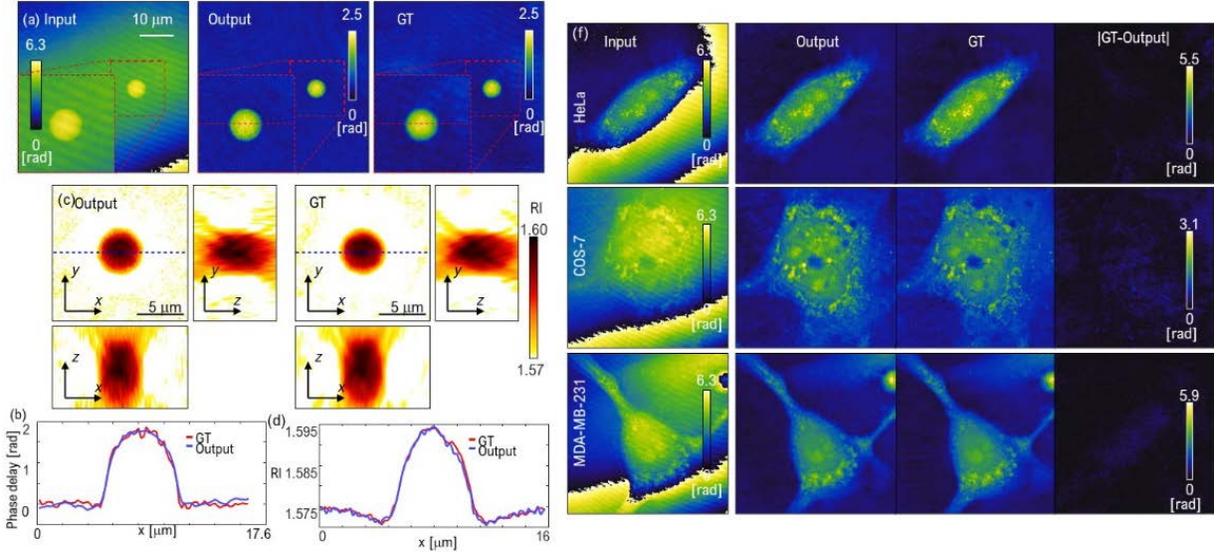

Fig. 5. The aberration of optical fields for the bead, which is a significantly different category of samples to the training dataset can be removed. (a) Phase images of a PS microsphere. The input phase image embracing aberration and phase wrapping (left), our deep learning result (middle), and the GT (right), obtained from the BGSM, are visualized. (b) Phase delay along the red dashed lines in the insets of (a) are plotted. (c) (left) The tomogram of the PS bead that is reconstructed from the aberration-removed fields in (a) using the proposed method. (right) The GT tomogram of the PS bead that is reconstructed using the BGSM. (d) RI values along the blue dashed lines in (c) are plotted. (f) Experimental results for three types of cells (HeLa, COS-8, and MDA-MB-231).

We also compared our method to the BGSM on the performance of optical diffraction tomography in Fig. 5(c). After 49 angles of optical fields were obtained, each set of optical fields were processed with our method and BGSM, respectively. The bead tomograms reconstructed from two different sets of optical fields are shown: our method (left) and BGSM (right). RI values along the blue dashed lines in Fig. 5(c) are plotted in Fig. 5(d). Although the network was trained using the dataset that consists of only the eukaryotic cells, our method has a close match with the BGSM in both 2-D QPI and tomographic reconstruction. Also, we have applied the present algorithm to various types of cancer cell lines [Fig. 5(f)]. The QPI images of the HeLa, COS-7, and MDA-MB-213 cell were clearly retrieved with the present method.

## Appendix E: Statistical analysis of results

The two types of errors (FCE and phase RMSE) are calculated for whole datasets (Fig. 6). For the images obtained from normal illumination only, 85% of the data in the test set has smaller FCE and phase RMSE than 0.028 and 0.15 rad, respectively. For the whole dataset containing images obtained from angled illumination, 85% of the data in the test set has smaller FCE and phase RMSE than 0.10 and 0.33 rad, respectively.

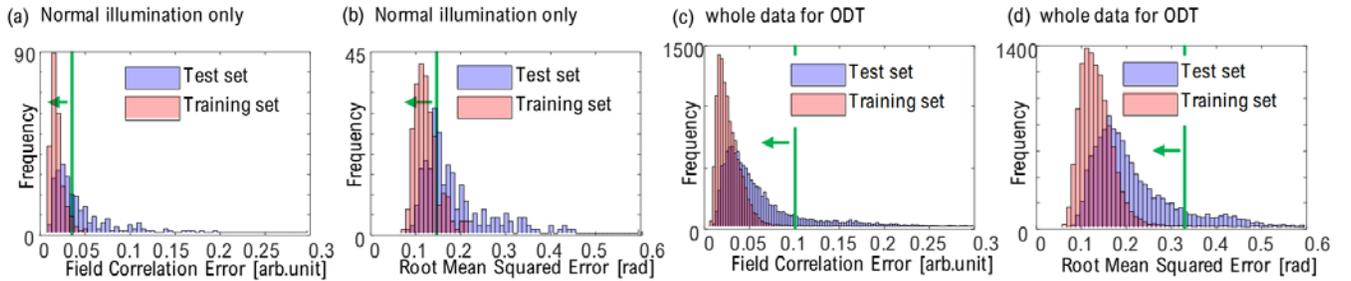

Fig. 6. The performance of the proposed method is quantitatively analyzed. The entire dataset consisting of the training set and test set is assessed by the errors between generated fields and the ground truth fields (fields obtained from background subtraction). The green lines indicate the threshold that contains 85% of the data in the test set.

## Funding

## Acknowledgment

The authors appreciate Dr. Weisun Park (KAIST) and Dr. Sumin Lee (Tomocube Inc.) for providing biological samples.

## Disclosures

Y. Jo, H.S. Min, G. Choi, and Y.K. Park have financial interests in Tomocube Inc., a company that commercializes ODT instruments and is one of the sponsors of the work.